\documentclass[]{pasj01}

\usepackage{mathpazo}
\usepackage[T1]{fontenc} 
\usepackage{lineno}
\usepackage{changes}
\definechangesauthor[name={Mitya}, color=orange]{DK}
\newcommand\Gas{\mathrm{gas}}
\newcommand\Mag{\mathrm{mag}}
\newcommand\EC{\mathrm{EC}}
\newcommand\Acc{\mathrm{acc}}
\newcommand\CR{\mathrm{cr}}

\Received{$\langle$reception date$\rangle$}
\Accepted{$\langle$acception date$\rangle$}
\Published{$\langle$publication date$\rangle$}

\begin{document}

\title{Upper Limit on the Coronal Cosmic Ray Energy Budget in Seyfert Galaxies}
\author{Yoshiyuki \textsc{Inoue}\altaffilmark{1,2,3}}
\altaffiltext{1}{Department of Earth and Space Science, Graduate School of Science, Osaka University, Toyonaka, Osaka 560-0043, Japan}
\altaffiltext{2}{Interdisciplinary Theoretical \& Mathematical Science Program (iTHEMS), RIKEN, 2-1 Hirosawa, Saitama 351-0198, Japan}
\altaffiltext{3}{Kavli Institute for the Physics and Mathematics of the Universe (WPI), UTIAS, The University of Tokyo, Kashiwa, Chiba 277-8583, Japan}
\email{yoshiyuki.inoue.sci@osaka-u.ac.jp}

\author{Shinsuke \textsc{Takasao}\altaffilmark{1}}
\email{takasao@astro-osaka.jp}

\author{Dmitry \textsc{Khangulyan}\altaffilmark{4, 5, 6}}
\altaffiltext{4}{Key Laboratory of Particle Astrophyics, Institute of High Energy Physics, Chinese Academy of Sciences, 100049 Beijing, China}
\altaffiltext{5}{Tianfu Cosmic Ray Research Center, 610000 Chengdu, Sichuan, China}
\altaffiltext{6}{Institute for Cosmic Ray Research, The University of Tokyo, 5-1-5 Kashiwa-no-Ha, Kashiwa City, Chiba, 277-8582, Japan}
\email{khangulyan@ihep.ac.cn}


\maketitle

\begin{abstract}
The IceCube collaboration has reported possible detections of high-energy neutrinos from nearby Seyfert galaxies. While central hot coronae are proposed as the primary neutrino production site, the exact coronal cosmic-ray energy budget has been loosely constrained. In this study, we propose a new stringent upper bound on the coronal cosmic-ray energy budget of Seyfert galaxies, considering both accretion dynamics and observed properties of radio-quiet Seyfert galaxies. Notably, even under the calorimetric condition where cosmic rays lose all their energies, our limit indicates that the coronal neutrino flux of NGC~1068 is about an order of magnitude fainter than the observed levels. This discrepancy suggests the need for further theoretical and observational investigations on the IceCube signals from Seyfert galaxies.
\end{abstract}

\section{Introduction}
The IceCube Collaboration has reported a 4.2-$\sigma$ detection of high-energy neutrinos from a nearby Seyfert galaxy, NGC~1068, based on 10 years of survey data \citep{IceCube2020PhRvL.124e1103A, IceCube2022icrc.confE1142G}, where Seyfert galaxies are a subset of active galactic nuclei (AGNs) characterized by a lack of pronounced jet activity. Subsequently, they have also identified 2.9-$\sigma$ signals from two other proximate Seyfert galaxies, NGC~4151 and CGCG~420-015 \citep{IceCube2023arXiv230715349G, IceCube2023arXiv230800024G}. Intriguingly, the neutrino flux of NGC~1068, as reported, exceeds its GeV gamma-ray flux \citep{IceCube2020PhRvL.124e1103A, IceCube2022Sci...378..538I, Ajello2023ApJ...954L..49A}, which implies a significant attenuation of GeV gamma rays. Although the exact neutrino fluxes of NGC~4151 and CGCG~420-015 have not been determined yet, this discrepancy would also apply to these two Seyferts considering the Fermi observations \citep{Fermi2022ApJS..260...53A, Peretti2023arXiv230303298P}.


This discrepancy between gamma rays and neutrinos has led to the proposal of various models to account for the neutrino signals from Seyferts, such as 
accretion-powered winds from supercritically accreting compact binaries \citep{Sridhar2022arXiv221211236S},
muon pair production in AGN core \citep{Hooper2023arXiv230506375H},
interaction between clouds and the accretion disk \citep{Muller2020A&A...636A..92M},
interaction between failed winds and the accretion disk \citep{Inoue2022arXiv220702097I},
jet interaction with interstellar medium \citep{Fang2023arXiv230707121F},
or stellar-mass black holes (BHs) embedded in the AGN accretion disk \citep{Tagawa2023ApJ...955...23T}. Here, given various known observational properties, the hot plasma surrounding the central supermassive black hole (SMBH) and its accretion disk, namely the corona, emerges as one of the most plausible sites for neutrino production \citep{Inoue2020ApJ...891L..33I, Murase2020PhRvL.125a1101M, Gutierrez2021A&A...649A..87G, Kheirandish2021ApJ...922...45K, Eichmann2022ApJ...939...43E, Murase2022ApJ...941L..17M, Neronov2023arXiv230609018N, Blanco2023arXiv230703259B, Ajello2023ApJ...954L..49A,
Mbarek2023arXiv231015222M,
Fiorillo2023arXiv231018254F}. 

The influence of high-energy cosmic rays (CRs) in SMBH coronae have been extensively investigated in the literature (see, e.g., \cite{Zdziarski1986, Kazanas1986, Sikora1987, Begelman1990, Stecker1992, Kalashev2015, Inoue2019ApJ...880...40I}). However, a definitive constraint on the cosmic-ray energy budget within these coronal regions remains elusive, primarily due to the ambiguous non-thermal activity of Seyferts in electromagnetic spectra. To account for the observed neutrino signals, a specific CR energy budget is often postulated without strict constraints. In this paper, we aim to constrain this CR energy budget by considering accretion dynamics and the observed properties of the coronae. Furthermore, we explore potential limits on the high-energy neutrino luminosities characteristic of Seyfert galaxies. 

\section{Physical Properties of AGN Coronae}
Accretion disk coronae serve as the primary sources of AGN X-ray emission through Comptonization of accretion disk photons \citep{Katz1976, 1977A&A....59..111B, Pozdniakov1977, Sunyaev1980}. Through X-ray spectral analyses, we can deduce thermal coronal parameters, specifically the coronal electron temperature $T_e$ and the Thomson scattering optical depth $\tau_\mathrm{X}$ \citep{Brenneman2014,Fabian2015, Ricci2018MNRAS.480.1819R}. X-ray AGNs typically have $k_B T_e\sim50$~keV, where $k_B$ is the Boltzmann constant,  and $\tau_\mathrm{X}\sim1$ \citep{Fabian2015, Ricci2018MNRAS.480.1819R, Pal2023arXiv231018196P}.

The coronal size $R_c$ is measured as $R_c\equiv r_cR_s\sim5$~--~$40~R_s$, where $R_s$ is the Schwarzschild radius, by optical--X-ray spectral fitting studies \citep{Jin2012, Kubota2018MNRAS.480.1247K}, micorolensing observations \citep{Morgan2012}, and coronal synchrotron emission studies \citep{Inoue2018ApJ...869..114I, Michiyama2023PASJ..tmp...58M}. In this paper, we adopt $r_c = 10$ as a fiducial value. This size is sufficient to attenuate GeV gamma-ray photons, which is required to match with the IceCube results of NGC~1068 \citep{Inoue2020ApJ...891L..33I, Murase2022ApJ...941L..17M}.

Theoretically, it is expected that BH coronae are parts of hot accretion flows, heated through advection \citep{Liu2002ApJ...575..117L, Kawanaka2008PASJ...60..399K, Kawabata2010PASJ...62..621K}, resulting in two-temperature plasma \citep{Narayan1994ApJ...428L..13N, Manmoto1997ApJ...489..791M}. Such two-temperature plasma has been also confirmed by recent numerical simulations (e.g., \cite{Liska2022ApJ...935L...1L}). Hereinafter, for the sake of simplicity, we treat coronae as uniform and steady two-temperature plasma surrounding BHs. 

The two-temperature plasma model naturally accounts for the observed coronal electron temperature in BH accretion systems (e.g., \cite{Kawabata2010PASJ...62..621K, Inoue2019ApJ...880...40I}). In the hot accretion flow, the ion temperature $T_i$ is close to the virial temperature \citep{Kato2008, Yuan2014}
\begin{equation}
\label{eq:t_ion}
T_i\approx \frac{GM_\mathrm{BH}m_p}{6k_B R_c}=\frac{m_pc^2}{12k_Br_c}
\simeq 9.1\times10^{10}~\mathrm{K}~\left(\frac{r_c}{10}\right)^{-1},
\end{equation}
where $G$ is the gravitational constant, $M_\mathrm{BH}$ is the BH mass, $m_p$ is the proton rest mass, and $c$ is the speed of light. This temperature translates to $k_BT_i\sim2.6~\mathrm{MeV}$. 

Through elastic Coulomb (EC) collisions, protons redistribute their heat to electrons, which are in turn cooled via Comptonization. The electron heating rate through EC collisions is given by
\begin{equation}
\frac{dT_e}{dt}\approx\frac{T_i}{t_{\EC,pe}}
\approx \frac{n_e\sigma_\mathrm{T} c\ln \Lambda}{\sqrt{\pi/2}} \left(\frac{m_e}{m_p}\right)T_i\theta_e^{-3/2},
\end{equation}
where $t_{\EC,pe}$ is the proton-electron Coulomb collision timescale \citep{Spitzer1962, Stepney1983}, $n_e$ is the coronal electron density, $\sigma_\mathrm{T}$ is the Thomson scattering cross section, $\ln \Lambda\approx20$ is the Coulomb logarithm, and $\theta_e \equiv k_BT_e/m_ec^2$ is the dimensionless electron temperature. We ignore the term of the dimensionless ion temperature $\theta_i\equiv k_BT_i/m_pc^2$ in $t_{\EC, pe}$, as $\theta_i\ll\theta_e$. 

The electron cooling rate due to Comptonization is described as
\begin{equation}
\frac{dT_e}{dt}\approx-\frac{4}{3}	\frac{\sigma_\mathrm{T} u_\mathrm{ph}T_e}{m_ec},
\end{equation}
where $u_\mathrm{ph}$ is the accretion disk photon energy density. We define the bolometric disk luminosity as $L_\mathrm{bol}=\epsilon L_\mathrm{Edd}$, where $\epsilon$ is the radiative efficiency and $L_\mathrm{Edd}$ is the Eddington luminosity. Thus, $u_\mathrm{ph} = L_\mathrm{bol}/4\pi R_c^2 c$.

By balancing these electron heating and cooling rates, the coronal electron temperature is estimated to be 
\begin{eqnarray}
    k_BT_e&\approx& \left(\frac{3\ln \Lambda}{4\sqrt{\pi/2}}\frac{m_e}{m_p}\frac{n_e}{u_\mathrm{ph}}k_BT_i\right)^{2/5}m_ec^2\\ \nonumber
&\simeq&84\left(\frac{\tau_\mathrm{X}}{1}\right)^{2/5}\left(\frac{\epsilon}{0.1}\right)^{-2/5}~[\rm keV],
\end{eqnarray}
where, given the coronal opacity and the size, the electron number density is expressed as $ n_e \approx {\tau_\mathrm{X}}/{\sigma_\mathrm{T}R_c}$.
The dependencies on $\tau_X$ and $\epsilon$ are retained. This temperature aligns closely with the measured coronal temperature $\sim50~\mathrm{keV}$ \citep{Fabian2015,Pal2023arXiv231018196P}. Therefore, the two-temperature plasma model provides a coherent explanation for the observed coronal temperatures.

\section{Bounds on Coronal CR Properties}
\subsection{Coronal CR Pressure}
The total pressure in the coronal flow $p$ is expressed as
\begin{equation}
p = p_\Gas + p_\Mag + p_\CR,
\end{equation}
where $p_\Gas$ is the gas pressure, $p_\Mag$ is the magnetic pressure, and $p_\CR$ is the cosmic-ray pressure. Here, we assume isotropic momentum distribution for CR particles. We neglect the radiation pressure because we consider the accretion rate regime $\dot{m}\ll1$, where $\dot{m}$ is the dimensionless mass accretion rate, scaled by the Eddington rate.

Given that the ion temperature $T_i$ exceeds the electron temperature $T_e$, the coronal thermal gas pressure is dominated by ions, which is
\begin{equation}
    \label{eq:p_gas_6}
    p_\Gas = n_i k_B T_i + n_e k_B T_e\approx n_i k_B T_i,
\end{equation}
where, assuming the charge neutrality, the ion number density $n_i$ is set equal to $n_e$. Then, we have the coronal gas pressure as 
\begin{equation}
    \label{eq:p_gas}
    p_\Gas\approx\frac{m_p c^4}{24G\sigma_\mathrm{T}}\frac{\tau_\mathrm{X}}{M_\mathrm{BH}r_c^2}.
\end{equation}
If we consider positron contribution, $n_i$ becomes smaller than $n_e$, so $p_\Gas$ would decrease. 

Magnetic fields accrete onto the central BH together with gas particles. The ratio of the gas pressure against the magnetic pressure is commonly defined by the plasma beta as
\begin{equation}
\beta\equiv \frac{p_\Gas}{p_\Mag},
\end{equation}
where we consider global coronal magnetic field.

In this paper, we introduce the ratio of the CR pressure against the magnetic pressure as
\begin{equation}
\label{eq:delta}
\delta \equiv \frac{p_\CR}{p_\Mag}.
\end{equation}
Observationally, as described above, coronal gamma-ray fluxes need to be attenuated by coronal X-ray photons. This implies that coronal CR particles should be generated and confined in the coronal region. Therefore, $\delta$, which limits the CR energy budget in the coronae, is bounded as
\begin{equation}
\delta \le 1.
\label{eq:cr_constraint}
\end{equation}
This condition is also applicable if CR particles are accelerated by magnetic activity associated with coronal flows as we consider the time-averaged structure. When this CR trapping condition reaches $\delta=1$, it signifies energy equipartition between magnetic fields and CR particles. It should be noted that in scenarios where $\delta > 1$ and $\beta < 1$, the CR pressure primarily sustains the accretion disk. This leads to distinct accretion dynamics compared to those dominated by gas, radiation, or magnetic pressure, which have been explored in literature.

Incorporating these considerations, the CR pressure is bounded in terms of the gas pressure as (Eqs.~\ref{eq:p_gas}--\ref{eq:delta})
\begin{equation}
\label{eq:p_cr}
    p_\CR = \frac{\delta}{\beta} p_\Gas \approx \frac{m_p c^4}{24G\sigma_\mathrm{T}}\frac{\delta}{\beta}\frac{\tau_\mathrm{X}}{M_\mathrm{BH}r_c^2}
\end{equation}

\subsection{Bounds on Coronal Cosmic Ray and Neutrino Emissions}
The energy density of cosmic rays, denoted as $u_\CR$, is associated with the CR pressure as $u_\CR = 3 p_\CR$. Those CRs generated in the coronal regions are transported to the further inner region while being trapped in turbulent magnetic fields in the accretion flow. As we consider geometrically thick coronae, we apply spherical approximation here. 

Since the energy source is the accretion process, the power of processes in the corona is limited by the rate at which the energy is supplied to the corona by the accretion flow. Using the standard expression for the energy flux (see, e.g., \citet{Landau1987Fluid}) one obtains
\begin{equation}
  \label{eq:L_CR}
    L_\CR \approx 4\pi R_c^2 v_\mathrm{acc} u_\CR,
\end{equation}
where $v_\Acc$ represents the accretion speed. Utilizing the self-similar solutions \citep{Narayan1994ApJ...428L..13N, Yuan2012ApJ...761..129Y, Yuan2014}, $v_\Acc$ is approximated $v_\Acc \simeq 1.1\times10^{10}\alpha r_c^{-1/2}~\mathrm{cm~s^{-1}}$,
where $\alpha$ is the dimensionless parameter representing the kinematic viscosity of the accretion flow, the so-called $\alpha$ parameter, which is typically set to be $0.1$. 

By combining Eqs.~\ref{eq:cr_constraint}, \ref{eq:p_cr} and \ref{eq:L_CR}, we can constrain the CR power as
\begin{eqnarray}
    L_\CR &\lesssim {7.3\times10^{41}}~\mathrm{erg~s^{-1}} 
    \left(\frac{\alpha}{0.1}\right)
    \left(\frac{\beta}{10}\right)^{-1}\nonumber \\
    &\times
    \left(\frac{\delta}{1}\right) 
    \left(\frac{M_\mathrm{BH}}{10^7M_\odot}\right)
    \left(\frac{r_c}{10}\right)^{-1/2}
    \left(\frac{\tau_X}{1}\right) \label{eq:L_CR_bound}
\end{eqnarray}
Here, if energy is injected into CRs from other than accretion processes (see e.g., \cite{Blandford2022MNRAS.514.5141B} for the ejection disk model), higher CR power might be realized.

The neutrino production channels are primarily $pp$ and $p\gamma$ interactions. One needs to consider the efficiency of $pp$ and $p\gamma$ processes. However, for the purpose of this study, we assume the calorimetric limit, where cosmic rays lose all their energies, for both processes to establish bounds on the neutrino luminosity to derive conservative upper limits. The bolometric neutrino luminosity per flavor is approximated as
\begin{equation}
L_\nu \approx f_\nu L_\CR \mathcal{C}^{-1},
\end{equation}
where $f_\nu$ is the neutrino production fraction. $f_\nu\simeq1/6$ for $pp$, while $f_\nu\simeq1/8$ for $p\gamma$. $\mathcal{C}$ reflects the spectral shape effect, which should be larger than unity. To derive the upper bound on the coronal neutrino luminosity, we take $f_\nu = 1/6$. Consequently, the coronal neutrino luminosity is bounded as
\begin{eqnarray}
    L_\nu &\le &{1.2}\times10^{41}~\mathrm{erg~s^{-1}} 
    \left(\frac{\alpha}{0.1}\right)
    \left(\frac{\beta}{10}\right)^{-1}\nonumber \\
    &\times&
    \left(\frac{\delta}{1}\right) 
    \left(\frac{\mathcal{C}}{1}\right)^{-1}
    \left(\frac{M_\mathrm{BH}}{10^7M_\odot}\right)
    \left(\frac{r_c}{10}\right)^{-1/2}
    \left(\frac{\tau_X}{1}\right). \label{eq:neutrino}
\end{eqnarray}

\subsection{Coronal Neutrino Emissions of NGC~1068, NGC~4151, and CGCG~420-015}
NGC~1068 is a Compton-thick Seyfert \citep{Bauer2015ApJ...812..116B, Marinucci2016MNRAS.456L..94M, Mizumoto2019ApJ...871..156M} located at the distance of $D_{\rm L}=13.97\pm2.10$~Mpc \citep{Anand2021MNRAS.501.3621A}. The mass of the central SMBH of NGC~1068 has been estimated using various methods. Water maser disk studies have reported a mass of $\sim10^7M_\odot$ \citep{Greenhill1996ApJ...472L..21G,Hure2002A&A...395L..21H,Lodato2003A&A...398..517L, Morishima2023PASJ...75...71M} \footnote{Studies focusing on the polarized broad Balmer emission line and the neutral FeK$\alpha$ line have indicated masses of $\sim7\times10^7M_\odot$ and $\sim1\times10^8M_\odot$, respectively \citep{Minezaki2015ApJ...802...98M}.}. We take $1\times10^7M_\odot$ for the SMBH mass of NGC~1068.

Employing Eq.~\ref{eq:neutrino}, we derive the upper bound for the total neutrino flux of
\begin{eqnarray}
    F_{\nu, \mathrm{NGC~1068}} &\le& {3.3\times}10^{-12}~\mathrm{TeV~cm^{-2}~s^{-1}}
    \nonumber\\
    &\times&\left(\frac{D_{\rm L}}{13.97~\mathrm{Mpc}}\right)^{-2}    \left(\frac{M_\mathrm{BH}}{10^7M_\odot}\right)
    \left(\frac{r_c}{10}\right)^{-1/2}
    \left(\frac{\tau_X}{1}\right) \nonumber\\ 
    &\times&\left(\frac{\alpha}{0.1}\right)
    \left(\frac{\beta}{10}\right)^{-1}
    \left(\frac{\delta}{1}\right) 
    \left(\frac{\mathcal{C}}{1}\right)^{-1}\,. \label{eq:ngc1068_flux}
\end{eqnarray}
If we assume all the neutrino flux is generated at 1~TeV, the resulting flux bound would be $3.3\times10^{-12}~\mathrm{TeV^{-1}~cm^{-2}~s^{-1}}$ at 1~TeV. Here, the observed neutrino flux of NGC~1068 is $\Phi_{\nu_{\mu}+\bar{\nu}_{\mu}}^{1\,\mathrm{TeV}}$\,=\,$(5.0 \pm 1.5_\mathrm{stat} \pm 0.6_\mathrm{sys}) \times 10^{-11}\,\mathrm{TeV^{-1}~cm^{-2}~s^{-1}}$ at 1~TeV with a spectral index of $3.2\pm0.2\pm0.07$ \citep{IceCube2022Sci...378..538I} which is about an order of magnitude higher than our bound. 

Similar disagreement is seen for the other two reported Seyferts. NGC~4151 is the X-ray brightest type-1 Seyfert \citep{Oh2018ApJS..235....4O} located at the distance of $D=15.8~\mathrm{Mpc}$ based on the Cepheid distance measurement \citep{Yuan2020ApJ...902...26Y}. Recently, GeV gamma-ray emission has been detected from this object \citep{Ajello2021ApJ...921..144A, Peretti2023arXiv230303298P}. NGC~4151 has a central SMBH mass of $2.1\times10^7~M_\odot$ based on reverberation mapping measurements \citep{DeRosa2018ApJ...866..133D}. CGCG~420-015 is a Compton-thick Seyfert \citep{Tanimoto2018ApJ...853..146T, Tanimoto2022ApJS..260...30T} at a distance of 127~Mpc \citep{Shimizu2016MNRAS.456.3335S}. The central SMBH mass is estimated as $2.0\times10^8~M_\odot$ \citep{Koss2017ApJ...850...74K}. The corresponding neutrino flux upper limit is $\le 5.3\times10^{-12}~\mathrm{TeV~cm^{-2}~s^{-1}}$ and $\le 7.9\times10^{-13}~\mathrm{TeV~cm^{-2}~s^{-1}}$ for NGC~4151 and CGCG~420-015, respectively. These upper limits are about an orders of magnitude lower than the $5\sigma$ sensitivity of IceCube \citep{IceCube2020PhRvL.124e1103A}.

These discrepancies in NGC~1068, NGC~4151, and CGCG~420-015 suggest potential contributions of other mechanisms, other compact sources, or potential background contamination. Lower $\beta$ would be able to reconcile the discrepancies. However, as described below, achieving low-beta plasma for the entire coronal region would not be straightforward, considering the nature of radio-quiet Seyferts and the dichotomy of AGNs.

\section{Discussion}
\subsection{Challenge of Strongly Magnetized Coronal Models}
When strong large-scale poloidal magnetic fields are present near the SMBHs, numerical simulations predict strong magnetic fields in the coronal region with values of $\beta\sim10^{-2}$--$10^{-3}$ \citep{Liska2022ApJ...935L...1L}. Such a low-$\beta$ accreting plasma would be necessary to successfully reproduce powerful jets seen in radio-loud AGNs. In these scenarios, AGN coronae could be magnetically heated by reconnections \citep{Haardt1991,Liu2002, Beloborodov2017}. However, our focus is on radio-quiet Seyfert galaxies, which, by definition, do not exhibit powerful jet activity. 


Lack of powerful jets in Seyferts implies that large-scale poloidal fields should not be dynamically important. In such cases, only weakly magnetized coronae will develop \citep{Liska2022ApJ...935L...1L}. Given the absence of strong, large-scale poloidal magnetic fields, the Parker instability \citep{Parker1955ApJ...121..491P, Parker1966ApJ...145..811P, Matsumoto1988PASJ...40..171M} could inhibit the magnetic field amplification, keeping $\beta\gtrsim10$ (see e.g., \cite{Takasao2018ApJ...857....4T, Hogg2018ApJ...861...24H, Dhang2019MNRAS.482..848D}). Consequently, this leads to  $\beta$ in the ranges of $\beta\sim10$--$100$.

We can relate the observed coronal magnetic field strength, $B_c$, to $\beta$, as the magnetic pressure is given by $p_\Mag\approx B^2/8\pi$. From observations, the coronal magnetic field $B_c$ of nearby Seyferts with SMBH masses of $10^7$--$10^8M_\odot$ is estimated to be in the range of $B_c\sim10$--$30~\mathrm{G}$ with $r_c\sim40$ \citep{Inoue2018ApJ...869..114I, Michiyama2023PASJ..tmp...58M}, leading to $\beta\approx100$. This estimate aligns with the arguments presented earlier. If this estimate holds, the coronal CR power could be reduced by an order of magnitude than our estimate, signifying the discrepancy between our bound and the observed neutrino flux.

The magnetic field estimates are based on the observations of millimeter (mm) emissions from Seyferts, utilizing the unprescended angular resolution and sensitivity of ALMA. These emissions are interpreted as coronal synchrotron emissions \citep{Inoue2018ApJ...869..114I, Michiyama2023PASJ..tmp...58M}. However, we note that the origin of these mm emissions remains a subject of debate (see, e.g., \cite{Raginski2016MNRAS.459.2082R, Baskin2021MNRAS.508..680B}). Nonetheless, the mm--X-ray luminosity correlation \citep{Kawamuro2022ApJ...938...87K, Ricci2023ApJ...952L..28R, Kawamuro2023arXiv230902776K} and the mm--X-ray time variability correlation \citep{Behar2020MNRAS.491.3523B, Chen2022MNRAS.515.1723C, Panessa2022MNRAS.510..718P, Petrucci2023arXiv230901804P} strongly suggest that both X-ray and mm emissions originate from the same compact coronal region.

Low-$\beta$ plasma might be prevalent in Seyfert coronae if the launch of jets is predominantly influenced by the angular momentum of the BHs. One widely accepted mechanism for the launch of powerful relativistic jets is the Blandford-Znajek process, in which the jet power is extracted by the rotation of SMBHs threaded by magnetic fields \citep{Blandford1977MNRAS.179..433B}. X-ray reflection spectra studies (e.g., \cite{Tanaka1995Natur.375..659T}) indicate that most AGNs including Seyferts harbor rapidly rotating SMBHs \citep{Vasudevan2016MNRAS.458.2012V, Reynolds2021ARA&A..59..117R}. This predominance of spinning SMBHs in AGNs is also theoretically anticipated, based on studies of accretion and merger histories of SMBHs \citep{Volonteri2013ApJ...775...94V}. Considering the insights from both theoretical and observational studies on SMBH spins, it is argued that a notable difference in magnetic flux strength, $\beta$, is inevitably necessary to explain the dichotomy of radio loudness of radio-loud and radio-quiet AGNs \citep{Sikora2007ApJ...658..815S, Sikora2013ApJ...764L..24S}. Consequently, the presence of low-$\beta$ coronal plasma in radio-quiet Seyferts would be unlikely, given their generally lower levels of jet activity compared to radio-loud populations.

 
\subsection{Particle Acceleration Processes}
Recent ALMA and IceCube observations indicate the presence of accelerated particles in AGN coronae \citep{Inoue2018ApJ...869..114I, Inoue2020ApJ...891L..33I, IceCube2022Sci...378..538I, Michiyama2023PASJ..tmp...58M}. Two-temperature accreting coronae are also characterized as collisionless plasma (e.g., \cite{Mahadevan1997ApJ...490..605M}), where non-thermal particles would be efficiently accelerated. Yet, the specific particle acceleration mechanisms at play within coronae remain elusive \citep{Inoue2019ApJ...880...40I, Murase2020PhRvL.125a1101M, Kheirandish2021ApJ...922...45K, Murase2022ApJ...941L..17M,Mbarek2023arXiv231015222M,Fiorillo2023arXiv231018254F}. 
\citet{Inoue2019ApJ...880...40I} have previously discussed potential acceleration processes in AGN coronae. We expand upon their arguments by combining our coronal CR-bound studies.

Firstly, diffusive shock acceleration (DSA) would be responsible for particle acceleration in coronae \citep{Stecker1992, Inoue2019ApJ...880...40I}. From the analysis of ALMA mm emission spectra, the injection spectral index of primary electrons at $\gamma_e\sim10$--$10^2$ is likely around $2$ \citep{Inoue2019ApJ...880...40I, Gutierrez2021A&A...649A..87G}. This injection index aligns with what is typically expected in a simple DSA scenario for a strong shock (e.g., \cite{Drury1983RPPh...46..973D, Blandford1987PhR...154....1B}). With $\beta\sim10$--$100$, DSA could potentially accelerate particles to energies of $>10$~TeV \citep{Inoue2019ApJ...880...40I}. Potential triggers for these shock accelerations include falling-back BLR clouds \citep{Muller2020A&A...636A..92M, Muller2022ApJ...931...39M, Sotomayor2022A&A...664A.178S} or failed-wind accretion  \citep{Inoue2022arXiv220702097I}. 

Secondly, stochastic acceleration might accelerate particles through the scattering by magnetic turbulence \citep{Kimura2019MNRAS.485..163K, Murase2020PhRvL.125a1101M, Eichmann2022ApJ...939...43E}. The stochastic acceleration is characterized by Alfv\'{e}n speed $v_A=B/\sqrt{4\pi m_p n_i}$, turbulence index $q$, and the ratio of the strength of turbulence against the background (e.g., \cite{Dermer1996}). The stochastic acceleration timescale depends on the magnetic field with $t_\mathrm{sta}\propto B^{-7/3}$ with the Kolomogorov index $q=5/3$ \citep{Inoue2019ApJ...880...40I, Eichmann2022ApJ...939...43E}. 
In low-$\beta$ plasma environments (i.e., in radio-loud AGNs), stochastic acceleration is likely to emerge as the predominant process for particle acceleration \citep{Zhdankin2018ApJ...867L..18Z, Zhdankin2019PhRvL.122e5101Z}. On the other hand, in radio-quiet AGNs having $\beta\sim10$--$100$, stochastic acceleration may not be as effective (but see also \cite{Kimura2015ApJ...806..159K, Lynn2014ApJ...791...71L, Sun2021MNRAS.506.1128S, Bacchini:2024oxs}).

Magnetic reconnection flares are also potential particle accelerators \citep{deGouveiadalPino2005A&A...441..845D, Singh2015ApJ...799L..20S,Kheirandish2021ApJ...922...45K,Fiorillo2023arXiv231018254F}. The specific particle acceleration mechanisms during these flares remain uncertain, with both DSA and stochastic acceleration being potential candidates (see e.g., \cite{Hoshino2012PhRvL.108m5003H, Nishizuka2013}). It would be beyond the scope of this paper to understand the specific acceleration process occurring in reconnection. However, we can estimate the achievable reconnection flare power. Differential rotation in accretion disks generates shear in magnetic fields, which would trigger magnetic reconnections. The achievable generation rate of magnetic energy in the coronal region can be estimated as 
\begin{equation}
P_\mathrm{mag}\approx \frac{\Omega_\mathrm{K}f_\mathrm{ani} B_\varphi^2R_c^3}{12\pi},
\end{equation}
where $\Omega_\mathrm{K}$ is the Keplerian angular velocity, $B_\varphi$ is the azimuthal magnetic field strength, and $f_\mathrm{ani}$ gives the ratio between radial and azimuthal components of magnetic field strengths. The rotation period rotation is $\approx2\pi/\Omega_\mathrm{K}\simeq1.7~\mathrm{days}(M_\mathrm{BH}/10^7M_\odot)\left({r_c}/{30}\right)^{3/2}$. Considering that a fraction of this magnetic energy $f_\mathrm{flare}$ will be released by reconnection \citep{Shibata2013PASJ...65...49S}, we estimate the total reconnection power as
\begin{eqnarray}
P_\mathrm{rec}&\approx&f_\mathrm{flare}P_\mathrm{mag}\\
    &\approx&f_\mathrm{flare}\frac{\Omega_\mathrm{K}f_\mathrm{ani} B_\varphi^2R_c^3}{12\pi}\\
    &\simeq&2.5\times10^{40}~\mathrm{erg~s^{-1}} 
    \left(\frac{f_\mathrm{flare}}{0.1}\right)
    \left(\frac{f_\mathrm{ani}}{1}\right)
    \left(\frac{\beta}{10}\right)^{-1} \nonumber \\
    &\times&\left(\frac{M_\mathrm{BH}}{10^7M_\odot}\right)
    \left(\frac{r_c}{10}\right)^{-1/2}
    \left(\frac{\tau_X}{1}\right) \label{Eq:Rec},
\end{eqnarray}
where we set $B_\varphi$ equal to the average $B$. This power is lower than our CR bound (Eq.~\ref{eq:L_CR_bound}).

As magnetic reconnections are transient phenomena and considering the radial dependence in Eq.~\ref{Eq:Rec}, we would expect sporadic giant reconnection flares. Recent numerical simulations found that such giant flare would indeed occur in the inner regions of accretion disks, even in globally high-$\beta$ plasma cases \citep{Takasao2019ApJ...878L..10T, Porth2021MNRAS.502.2023P, Ripperda2022ApJ...924L..32R, Ball2018ApJ...853..184B}. These reconnections could serve as a source of cosmic-ray particles or provide pre-accelerated particles. Detecting flaring variabilities with ALMA or IceCube would be crucial in confirming these phenomena. 

\subsection{Influence of Coronal Structure and Strong Gravity}
In our analysis, we have not explored the inner structure of the corona or the influence of gravity. By adopting a density structure from hot accretion flow without any outflows, $n_i\propto r^{-3/2}$ \citep{Yuan2014}, we estimate that the CR power could increase by a factor of 10 in the case of a maximally rotating BH. However, the actual inner structure of the corona diverges from this simple hot accretion flow model. For example, a more uniform density within $5$--$10 R_s$ is presented by a recent numerical study (e.g., \cite{Liska2022ApJ...935L...1L}).

Considering the coronae's proximity to the central BH, the impact of general relativity, especially gravitational redshift, is significant. This effect can dramatically reduce the observable flux from the core region. For rapidly spinning black holes, \citet{Hagen2023MNRAS.525.3455H} demonstrate that emissions from the central region could be reduced by over 80\% based on general relativistic ray-tracing studies. Thus, if neutrinos are produced deep within the coronae, the actual CR energy density would be considerably higher than currently estimated.

\section{Summary}
In this study, we explore the CR energy budget within the coronae of Seyfert galaxies. Taking into account the accretion dynamics and observed characteristics of Seyfert coronae, we establish a stringent upper limit on the coronal CR power, as denoted by Eq.~\ref{eq:L_CR_bound}. This constraint is more restrictive than those derived from bolometric luminosities. When considering the calorimetric limit, our established bound yields the neutrino flux that is approximately an order of magnitude lower than observations from the direction of NGC~1068. Such a discrepancy implies potential contributions from other compact regions or the possibility of background event contamination. A more in-depth combination of theoretical and observational analyses of the IceCube signals emanating from Seyfert galaxies is warranted.

Recent numerical simulations indicate a preference for high $\beta$ plasma ($\beta\gg1$) in AGNs lacking powerful jet activity \citep{Liska2022ApJ...935L...1L}. In such a scenario, the DSA process emerges as the favored particle acceleration mechanism. With a high $\beta$ value, both stochastic acceleration and magnetic reconnection flares face challenges in efficiently accelerating particles. Nonetheless, sporadic giant magnetic reconnection flares (e.g., \cite{Takasao2019ApJ...878L..10T, Porth2021MNRAS.502.2023P, Ripperda2022ApJ...924L..32R}), whose magnetic fields are locally amplified, would potentially supply the necessary cosmic-ray population.

\begin{ack}
The authors would like to thank Chris Done, Scott, Hagen, and Ellis Owen for useful comments and discussions. YI is supported by NAOJ ALMA Scientific Research Grant Number 2021-17A and JSPS KAKENHI Grant Number JP18H05458, JP19K14772, and JP22K18277. ST is supported by JSPS KAKENHI Grant Number JP21H04487, JP22KK0043, and JP22K14074. This work was supported by World Premier International Research Center Initiative (WPI), MEXT, Japan.  
\end{ack}


\begin{thebibliography}{122}
\expandafter\ifx\csname natexlab\endcsname\relax\def\natexlab#1{#1}\fi

\bibitem[{{Aartsen} {et~al.}(2020){Aartsen}, {Ackermann}, {Adams}, {Aguilar}, {Ahlers}, {Ahrens}, {Alispach}, {Andeen}, {Anderson}, {Ansseau}, {Anton}, {Arg{\"u}elles}, {Auffenberg}, {Axani}, {Backes}, {Bagherpour}, {Bai}, {Balagopal}, {Barbano}, {Barwick}, {Bastian}, {Baum}, {Baur}, {Bay}, {Beatty}, {Becker}, {Becker Tjus}, {BenZvi}, {Berley}, {Bernardini}, {Besson}, {Binder}, {Bindig}, {Blaufuss}, {Blot}, {Bohm}, {B{\"o}rner}, {B{\"o}ser}, {Botner}, {B{\"o}ttcher}, {Bourbeau}, {Bourbeau}, {Bradascio}, {Braun}, {Bron}, {Brostean-Kaiser}, {Burgman}, {Buscher}, {Busse}, {Carver}, {Chen}, {Cheung}, {Chirkin}, {Choi}, {Clark}, {Classen}, {Coleman}, {Collin}, {Conrad}, {Coppin}, {Correa}, {Cowen}, {Cross}, {Dave}, {De Clercq}, {DeLaunay}, {Dembinski}, {Deoskar}, {De Ridder}, {Desiati}, {de Vries}, {de Wasseige}, {de With}, {DeYoung}, {Diaz}, {D{\'\i}az-V{\'e}lez}, {Dujmovic}, {Dunkman}, {Dvorak}, {Eberhardt}, {Ehrhardt}, {Eller}, {Engel}, {Evenson}, {Fahey}, {Fazely}, {Felde}, {Filimonov}, {Finley}, {Fox},
  {Franckowiak}, {Friedman}, {Fritz}, {Gaisser}, {Gallagher}, {Ganster}, {Garrappa}, {Gerhardt}, {Ghorbani}, {Glauch}, {Gl{\"u}senkamp}, {Goldschmidt}, {Gonzalez}, {Grant}, {Griffith}, {Griswold}, {G{\"u}nder}, {G{\"u}nd{\"u}z}, {Haack}, {Hallgren}, {Halliday}, {Halve}, {Halzen}, {Hanson}, {Haungs}, {Hebecker}, {Heereman}, {Heix}, {Helbing}, {Hellauer}, {Henningsen}, {Hickford}, {Hignight}, {Hill}, {Hoffman}, {Hoffmann}, {Hoinka}, {Hokanson-Fasig}, {Hoshina}, {Huang}, {Huber}, {Huber}, {Hultqvist}, {H{\"u}nnefeld}, {Hussain}, {In}, {Iovine}, {Ishihara}, {Japaridze}, {Jeong}, {Jero}, {Jones}, {Jonske}, {Joppe}, {Kang}, {Kang}, {Kappes}, {Kappesser}, {Karg}, {Karl}, {Karle}, {Katz}, {Kauer}, {Kelley}, {Kheirandish}, {Kim}, {Kintscher}, {Kiryluk}, {Kittler}, {Klein}, {Koirala}, {Kolanoski}, {K{\"o}pke}, {Kopper}, {Kopper}, {Koskinen}, {Kowalski}, {Krings}, {Kr{\"u}ckl}, {Kulacz}, {Kurahashi}, {Kyriacou}, {Labare}, {Lanfranchi}, {Larson}, {Lauber}, {Lazar}, {Leonard}, {Leszczy{\'n}ska}, {Leuermann}, {Liu},
  {Lohfink}, {Lozano Mariscal}, {Lu}, {Lucarelli}, {L{\"u}nemann}, {Luszczak}, {Lyu}, {Ma}, {Madsen}, {Maggi}, {Mahn}, {Makino}, {Mallik}, {Mallot}, {Mancina}, {Mari{\c{s}}}, {Maruyama}, {Mase}, {Matis}, {Maunu}, {McNally}, {Meagher}, {Medici}, {Medina}, {Meier}, {Meighen-Berger}, {Menne}, {Merino}, {Meures}, {Micallef}, {Mockler}, {Moment{\'e}}, {Montaruli}, {Moore}, {Morse}, {Moulai}, {Muth}, {Nagai}, {Naumann}, {Neer}, {Niederhausen}, {Nisa}, {Nowicki}, {Nygren}, {Obertacke Pollmann}, {Oehler}, {Olivas}, {O'Murchadha}, {O'Sullivan}, {Palczewski}, {Pandya}, {Pankova}, {Park}, {Peiffer}, {P{\'e}rez de los Heros}, {Philippen}, {Pieloth}, {Pinat}, {Pizzuto}, {Plum}, {Porcelli}, {Price}, {Przybylski}, {Raab}, {Raissi}, {Rameez}, {Rauch}, {Rawlins}, {Rea}, {Reimann}, {Relethford}, {Renschler}, {Renzi}, {Resconi}, {Rhode}, {Richman}, {Robertson}, {Rongen}, {Rott}, {Ruhe}, {Ryckbosch}, {Rysewyk}, {Safa}, {Sanchez Herrera}, {Sandrock}, {Sandroos}, {Santander}, {Sarkar}, {Sarkar}, {Satalecka}, {Schaufel},
  {Schieler}, {Schlunder}, {Schmidt}, {Schneider}, {Schneider}, {Schr{\"o}der}, {Schumacher}, {Sclafani}, {Seckel}, {Seunarine}, {Shefali}, {Silva}, {Snihur}, {Soedingrekso}, {Soldin}, {Song}, {Spiczak}, {Spiering}, {Stachurska}, {Stamatikos}, {Stanev}, {Stein}, {Steinm{\"u}ller}, {Stettner}, {Steuer}, {Stezelberger}, {Stokstad}, {St{\"o}{\ss}l}, {Strotjohann}, {St{\"u}rwald}, {Stuttard}, {Sullivan}, {Taboada}, {Tenholt}, {Ter-Antonyan}, {Terliuk}, {Tilav}, {Tollefson}, {Tomankova}, {T{\"o}nnis}, {Toscano}, {Tosi}, {Trettin}, {Tselengidou}, {Tung}, {Turcati}, {Turcotte}, {Turley}, {Ty}, {Unger}, {Unland Elorrieta}, {Usner}, {Vandenbroucke}, {Van Driessche}, {van Eijk}, {van Eijndhoven}, {Vanheule}, {van Santen}, {Vraeghe}, {Walck}, {Wallace}, {Wallraff}, {Wandkowsky}, {Watson}, {Weaver}, {Weindl}, {Weiss}, {Weldert}, {Wendt}, {Werthebach}, {Whelan}, {Whitehorn}, {Wiebe}, {Wiebusch}, {Wille}, {Williams}, {Wills}, {Wolf}, {Wood}, {Wood}, {Woschnagg}, {Wrede}, {Xu}, {Xu}, {Xu}, {Yanez}, {Yodh}, {Yoshida},
  {Yuan}, \& {Z{\"o}cklein}}]{IceCube2020PhRvL.124e1103A}
{Aartsen}, M.~G., {Ackermann}, M., {Adams}, J., {et~al.} 2020, \prl, 124, 051103

\bibitem[{{Abdollahi} {et~al.}(2022){Abdollahi}, {Acero}, {Baldini}, {Ballet}, {Bastieri}, {Bellazzini}, {Berenji}, {Berretta}, {Bissaldi}, {Blandford}, {Bloom}, {Bonino}, {Brill}, {Britto}, {Bruel}, {Burnett}, {Buson}, {Cameron}, {Caputo}, {Caraveo}, {Castro}, {Chaty}, {Cheung}, {Chiaro}, {Cibrario}, {Ciprini}, {Coronado-Bl{\'a}zquez}, {Crnogorcevic}, {Cutini}, {D'Ammando}, {De Gaetano}, {Digel}, {Di Lalla}, {Dirirsa}, {Di Venere}, {Dom{\'\i}nguez}, {Fallah Ramazani}, {Fegan}, {Ferrara}, {Fiori}, {Fleischhack}, {Franckowiak}, {Fukazawa}, {Funk}, {Fusco}, {Galanti}, {Gammaldi}, {Gargano}, {Garrappa}, {Gasparrini}, {Giacchino}, {Giglietto}, {Giordano}, {Giroletti}, {Glanzman}, {Green}, {Grenier}, {Grondin}, {Guillemot}, {Guiriec}, {Gustafsson}, {Harding}, {Hays}, {Hewitt}, {Horan}, {Hou}, {J{\'o}hannesson}, {Karwin}, {Kayanoki}, {Kerr}, {Kuss}, {Landriu}, {Larsson}, {Latronico}, {Lemoine-Goumard}, {Li}, {Liodakis}, {Longo}, {Loparco}, {Lott}, {Lubrano}, {Maldera}, {Malyshev}, {Manfreda}, {Mart{\'\i}-Devesa},
  {Mazziotta}, {Mereu}, {Meyer}, {Michelson}, {Mirabal}, {Mitthumsiri}, {Mizuno}, {Moiseev}, {Monzani}, {Morselli}, {Moskalenko}, {Negro}, {Nuss}, {Omodei}, {Orienti}, {Orlando}, {Paneque}, {Pei}, {Perkins}, {Persic}, {Pesce-Rollins}, {Petrosian}, {Pillera}, {Poon}, {Porter}, {Principe}, {Rain{\`o}}, {Rando}, {Rani}, {Razzano}, {Razzaque}, {Reimer}, {Reimer}, {Reposeur}, {S{\'a}nchez-Conde}, {Saz Parkinson}, {Scotton}, {Serini}, {Sgr{\`o}}, {Siskind}, {Smith}, {Spandre}, {Spinelli}, {Sueoka}, {Suson}, {Tajima}, {Tak}, {Thayer}, {Thompson}, {Torres}, {Troja}, {Valverde}, {Wood}, \& {Zaharijas}}]{Fermi2022ApJS..260...53A}
{Abdollahi}, S., {Acero}, F., {Baldini}, L., {et~al.} 2022, \apjs, 260, 53

\bibitem[{{Ajello} {et~al.}(2021){Ajello}, {Baldini}, {Ballet}, {Barbiellini}, {Bastieri}, {Bellazzini}, {Berretta}, {Bissaldi}, {Blandford}, {Bloom}, {Bonino}, {Bruel}, {Buson}, {Cameron}, {Caprioli}, {Caputo}, {Cavazzuti}, {Chartas}, {Chen}, {Cheung}, {Chiaro}, {Costantin}, {Cutini}, {D'Ammando}, {de la Torre Luque}, {de Palma}, {Desai}, {Diesing}, {Di Lalla}, {Dirirsa}, {Di Venere}, {Dom{\'\i}nguez}, {Fegan}, {Franckowiak}, {Fukazawa}, {Funk}, {Fusco}, {Gargano}, {Gasparrini}, {Giglietto}, {Giordano}, {Giroletti}, {Green}, {Grenier}, {Guiriec}, {Hartmann}, {Horan}, {J{\'o}hannesson}, {Karwin}, {Kerr}, {Kova{\v{c}}evi{\'c}}, {Kuss}, {Larsson}, {Latronico}, {Lemoine-Goumard}, {Li}, {Liodakis}, {Longo}, {Loparco}, {Lovellette}, {Lubrano}, {Maldera}, {Manfreda}, {Marchesi}, {Marcotulli}, {Mart{\'\i}-Devesa}, {Mazziotta}, {Mereu}, {Michelson}, {Mizuno}, {Monzani}, {Morselli}, {Moskalenko}, {Negro}, {Omodei}, {Orienti}, {Orlando}, {Paliya}, {Paneque}, {Pei}, {Persic}, {Pesce-Rollins}, {Porter}, {Principe},
  {Racusin}, {Rain{\`o}}, {Rando}, {Rani}, {Razzano}, {Reimer}, {Reimer}, {Saz Parkinson}, {Serini}, {Sgr{\`o}}, {Siskind}, {Spandre}, {Spinelli}, {Suson}, {Tak}, {Torres}, {Troja}, {Wood}, {Zaharijas}, \& {Zrake}}]{Ajello2021ApJ...921..144A}
{Ajello}, M., {Baldini}, L., {Ballet}, J., {et~al.} 2021, \apj, 921, 144

\bibitem[{{Ajello} {et~al.}(2023){Ajello}, {Murase}, \& {McDaniel}}]{Ajello2023ApJ...954L..49A}
{Ajello}, M., {Murase}, K., \& {McDaniel}, A. 2023, \apjl, 954, L49

\bibitem[{{Anand} {et~al.}(2021){Anand}, {Lee}, {Van Dyk}, {Leroy}, {Rosolowsky}, {Schinnerer}, {Larson}, {Kourkchi}, {Kreckel}, {Scheuermann}, {Rizzi}, {Thilker}, {Tully}, {Bigiel}, {Blanc}, {Boquien}, {Chandar}, {Dale}, {Emsellem}, {Deger}, {Glover}, {Grasha}, {Groves}, {S. Klessen}, {Kruijssen}, {Querejeta}, {S{\'a}nchez-Bl{\'a}zquez}, {Schruba}, {Turner}, {Ubeda}, {Williams}, \& {Whitmore}}]{Anand2021MNRAS.501.3621A}
{Anand}, G.~S., {Lee}, J.~C., {Van Dyk}, S.~D., {et~al.} 2021, \mnras, 501, 3621

\bibitem[{Bacchini {et~al.}(2024)Bacchini, Zhdankin, Gorbunov, Werner, Arzamasskiy, Begelman, \& Uzdensky}]{Bacchini:2024oxs}
Bacchini, F., Zhdankin, V., Gorbunov, E.~A., {et~al.} 2024, arXiv e-prints, arXiv:2401.01399

\bibitem[{{Ball} {et~al.}(2018){Ball}, {{\"O}zel}, {Psaltis}, {Chan}, \& {Sironi}}]{Ball2018ApJ...853..184B}
{Ball}, D., {{\"O}zel}, F., {Psaltis}, D., {Chan}, C.-K., \& {Sironi}, L. 2018, \apj, 853, 184

\bibitem[{{Baskin} \& {Laor}(2021)}]{Baskin2021MNRAS.508..680B}
{Baskin}, A. \& {Laor}, A. 2021, \mnras, 508, 680

\bibitem[{{Bauer} {et~al.}(2015){Bauer}, {Ar{\'e}valo}, {Walton}, {Koss}, {Puccetti}, {Gandhi}, {Stern}, {Alexander}, {Balokovi{\'c}}, {Boggs}, {Brandt}, {Brightman}, {Christensen}, {Comastri}, {Craig}, {Del Moro}, {Hailey}, {Harrison}, {Hickox}, {Luo}, {Markwardt}, {Marinucci}, {Matt}, {Rigby}, {Rivers}, {Saez}, {Treister}, {Urry}, \& {Zhang}}]{Bauer2015ApJ...812..116B}
{Bauer}, F.~E., {Ar{\'e}valo}, P., {Walton}, D.~J., {et~al.} 2015, \apj, 812, 116

\bibitem[{{Begelman} {et~al.}(1990){Begelman}, {Rudak}, \& {Sikora}}]{Begelman1990}
{Begelman}, M.~C., {Rudak}, B., \& {Sikora}, M. 1990, \apj, 362, 38

\bibitem[{{Behar} {et~al.}(2020){Behar}, {Kaspi}, {Paubert}, {Billot}, {Peretz}, {Baldi}, {Laor}, {Kaastra}, \& {Mehdipour}}]{Behar2020MNRAS.491.3523B}
{Behar}, E., {Kaspi}, S., {Paubert}, G., {et~al.} 2020, \mnras, 491, 3523

\bibitem[{{Beloborodov}(2017)}]{Beloborodov2017}
{Beloborodov}, A.~M. 2017, \apj, 850, 141

\bibitem[{{Bisnovatyi-Kogan} \& {Blinnikov}(1977)}]{1977A&A....59..111B}
{Bisnovatyi-Kogan}, G.~S. \& {Blinnikov}, S.~I. 1977, \aap, 59, 111

\bibitem[{{Blanco} {et~al.}(2023){Blanco}, {Hooper}, {Linden}, \& {Pinetti}}]{Blanco2023arXiv230703259B}
{Blanco}, C., {Hooper}, D., {Linden}, T., \& {Pinetti}, E. 2023, arXiv e-prints, arXiv:2307.03259

\bibitem[{{Blandford} \& {Eichler}(1987)}]{Blandford1987PhR...154....1B}
{Blandford}, R. \& {Eichler}, D. 1987, \physrep, 154, 1

\bibitem[{{Blandford} \& {Globus}(2022)}]{Blandford2022MNRAS.514.5141B}
{Blandford}, R. \& {Globus}, N. 2022, \mnras, 514, 5141

\bibitem[{{Blandford} \& {Znajek}(1977)}]{Blandford1977MNRAS.179..433B}
{Blandford}, R.~D. \& {Znajek}, R.~L. 1977, \mnras, 179, 433

\bibitem[{{Brenneman} {et~al.}(2014){Brenneman}, {Madejski}, {Fuerst}, {Matt}, {Elvis}, {Harrison}, {Ballantyne}, {Boggs}, {Christensen}, {Craig}, {Fabian}, {Grefenstette}, {Hailey}, {Madsen}, {Marinucci}, {Rivers}, {Stern}, {Walton}, \& {Zhang}}]{Brenneman2014}
{Brenneman}, L.~W., {Madejski}, G., {Fuerst}, F., {et~al.} 2014, \apj, 788, 61

\bibitem[{{Chen} {et~al.}(2022){Chen}, {Laor}, \& {Behar}}]{Chen2022MNRAS.515.1723C}
{Chen}, S., {Laor}, A., \& {Behar}, E. 2022, \mnras, 515, 1723

\bibitem[{{de Gouveia dal Pino} \& {Lazarian}(2005)}]{deGouveiadalPino2005A&A...441..845D}
{de Gouveia dal Pino}, E.~M. \& {Lazarian}, A. 2005, \aap, 441, 845

\bibitem[{{De Rosa} {et~al.}(2018){De Rosa}, {Fausnaugh}, {Grier}, {Peterson}, {Denney}, {Horne}, {Bentz}, {Ciroi}, {Dalla Bont{\`a}}, {Joner}, {Kaspi}, {Kochanek}, {Pogge}, {Sergeev}, {Vestergaard}, {Adams}, {Antognini}, {Araya Salvo}, {Armstrong}, {Bae}, {Barth}, {Beatty}, {Bhattacharjee}, {Borman}, {Boroson}, {Bottorff}, {Brown}, {Brown}, {Brotherton}, {Coker}, {Clanton}, {Cracco}, {Crawford}, {Croxall}, {Eftekharzadeh}, {Eracleous}, {Fiorenza}, {Frassati}, {Hawkins}, {Henderson}, {Holoien}, {Hutchison}, {Kellar}, {Kilerci-Eser}, {Kim}, {King}, {La Mura}, {Laney}, {Li}, {Lochhaas}, {Ma}, {MacInnis}, {Manne-Nicholas}, {Mason}, {McGraw}, {Mogren}, {Montouri}, {Moody}, {Mosquera}, {Mudd}, {Musso}, {Nazarov}, {Nguyen}, {Ochner}, {Okhmat}, {Onken}, {Ou-Yang}, {Pancoast}, {Pei}, {Penny}, {Poleski}, {Portaluri}, {Prieto}, {Price-Whelan}, {Pulatova}, {Rafter}, {Roettenbacher}, {Romero-Colmenero}, {Runnoe}, {Schimoia}, {Shappee}, {Sherf}, {Simonian}, {Siviero}, {Skowron}, {Skowron}, {Somers}, {Spencer}, {Starkey},
  {Stevens}, {Stoll}, {Tamajo}, {Tayar}, {van Saders}, {Valenti}, {Villanueva}, {Villforth}, {Weiss}, {Winkler}, {Zastrow}, {Zhu}, \& {Zu}}]{DeRosa2018ApJ...866..133D}
{De Rosa}, G., {Fausnaugh}, M.~M., {Grier}, C.~J., {et~al.} 2018, \apj, 866, 133

\bibitem[{{Dermer} {et~al.}(1996){Dermer}, {Miller}, \& {Li}}]{Dermer1996}
{Dermer}, C.~D., {Miller}, J.~A., \& {Li}, H. 1996, \apj, 456, 106

\bibitem[{{Dhang} \& {Sharma}(2019)}]{Dhang2019MNRAS.482..848D}
{Dhang}, P. \& {Sharma}, P. 2019, \mnras, 482, 848

\bibitem[{{Drury}(1983)}]{Drury1983RPPh...46..973D}
{Drury}, L.~O. 1983, Reports on Progress in Physics, 46, 973

\bibitem[{{Eichmann} {et~al.}(2022){Eichmann}, {Oikonomou}, {Salvatore}, {Dettmar}, \& {Tjus}}]{Eichmann2022ApJ...939...43E}
{Eichmann}, B., {Oikonomou}, F., {Salvatore}, S., {Dettmar}, R.-J., \& {Tjus}, J.~B. 2022, \apj, 939, 43

\bibitem[{{Fabian} {et~al.}(2015){Fabian}, {Lohfink}, {Kara}, {Parker}, {Vasudevan}, \& {Reynolds}}]{Fabian2015}
{Fabian}, A.~C., {Lohfink}, A., {Kara}, E., {et~al.} 2015, \mnras, 451, 4375

\bibitem[{{Fang} {et~al.}(2023){Fang}, {Lopez Rodriguez}, {Halzen}, \& {Gallagher}}]{Fang2023arXiv230707121F}
{Fang}, K., {Lopez Rodriguez}, E., {Halzen}, F., \& {Gallagher}, J.~S. 2023, arXiv e-prints, arXiv:2307.07121

\bibitem[{{Fiorillo} {et~al.}(2023){Fiorillo}, {Petropoulou}, {Comisso}, {Peretti}, \& {Sironi}}]{Fiorillo2023arXiv231018254F}
{Fiorillo}, D. F.~G., {Petropoulou}, M., {Comisso}, L., {Peretti}, E., \& {Sironi}, L. 2023, arXiv e-prints, arXiv:2310.18254

\bibitem[{{Glauch} {et~al.}(2023){Glauch}, {Kheirandish}, {Kontrimas}, {Liu}, \& {Niederhausen}}]{IceCube2023arXiv230800024G}
{Glauch}, T., {Kheirandish}, A., {Kontrimas}, T., {Liu}, Q., \& {Niederhausen}, H. 2023, arXiv e-prints, arXiv:2308.00024

\bibitem[{{Goswami}(2023)}]{IceCube2023arXiv230715349G}
{Goswami}, S. 2023, arXiv e-prints, arXiv:2307.15349

\bibitem[{{Goswami} {et~al.}(2022){Goswami}, {Privon}, {Santander}, {Abbasi}, {Ackermann}, {Adams}, {Aguilar}, {Ahlers}, {Ahrens}, {Alispach}, {Alves Junior}, {Amin}, {An}, {Andeen}, {Anderson}, {Anton}, {Arguelles}, {Ashida}, {Axani}, {Bai}, {Balagopal V.}, {Barbano}, {Barwick}, {Bastian}, {Basu}, {Baur}, {Bay}, {Beatty}, {Becker}, {Becker Tjus}, {Bellenghi}, {BenZvi}, {Berley}, {Bernardini}, {Besson}, {Binder}, {Bindig}, {Blaufuss}, {Blot}, {Boddenberg}, {Bontempo}, {Borowka}, {Boser}, {Botner}, {Bottcher}, {Bourbeau}, {Bradascio}, {Braun}, {Bron}, {Brostean-Kaiser}, {Browne}, {Burgman}, {Burley}, {Busse}, {Campana}, {Carnie-Bronca}, {Chen}, {Chirkin}, {Choi}, {Clark}, {Clark}, {Classen}, {Coleman}, {Collin}, {Conrad}, {Coppin}, {Correa}, {Cowen}, {Cross}, {Dappen}, {Dave}, {De Clercq}, {DeLaunay}, {Dembinski}, {Deoskar}, {De Ridder}, {Desai}, {Desiati}, {de Vries}, {de Wasseige}, {De With}, {DeYoung}, {Dharani}, {Diaz}, {Diaz-Velez}, {Dittmer}, {Dujmovic}, {Dunkman}, {DuVernois}, {Dvorak}, {Ehrhardt},
  {Eller}, {Engel}, {Erpenbeck}, {Evans}, {Evenson}, {Fan}, {Fazely}, {Fiedlschuster}, {Fienberg}, {Filimonov}, {Finley}, {Fischer}, {Fox}, {Franckowiak}, {Friedman}, {Fritz}, {Furst}, {Gaisser}, {Gallagher}, {Ganster}, {Garcia}, {Garrappa}, {Gerhardt}, {Ghadimi}, {Glaser}, {Glauch}, {Glusenkamp}, {Goldschmidt}, {Gonzalez}, {Grant}, {Gr{\'e}goire}, {Griswold}, {Gunduz}, {G{\"u}nther}, {Haack}, {Hallgren}, {Halliday}, {Halve}, {Halzen}, {Minh}, {Hanson}, {Hardin}, {Harnisch}, {Haungs}, {Hauser}, {Hebecker}, {Helbing}, {Henningsen}, {Hettinger}, {Hickford}, {Hignight}, {Hill}, {Hill}, {Hoffman}, {Hoffmann}, {Hoinka}, {Hokanson-Fasig}, {Hoshina}, {Huang}, {Huber}, {Huber}, {Hultqvist}, {Hunnefeld}, {Hussain}, {In}, {Iovine}, {Ishihara}, {Jansson}, {Japaridze}, {Jeong}, {Jones}, {Kang}, {Kang}, {Kang}, {Kappes}, {Kappesser}, {Karg}, {Karl}, {Karle}, {Katz}, {Kauer}, {Kellermann}, {Kelley}, {Kheirandish}, {Kin}, {Kintscher}, {Kiryluk}, {Klein}, {Koirala}, {Kolanoski}, {Kontrimas}, {Kopke}, {Kopper}, {Kopper},
  {Koskinen}, {Koundal}, {Kovacevich}, {Kowalski}, {Kozynets}, {Kun}, {Kurahashi}, {Lad}, {Lagunas Gualda}, {Lanfranchi}, {Larson}, {Lauber}, {Lazar}, {Lee}, {Leonard}, {Leszczy{\'n}ska}, {Li}, {Lincetto}, {Liu}, {Liubarska}, {Lohfink}, {Lozano Mariscal}, {Lu}, {Lucarelli}, {Ludwig}, {Luszczak}, {Lyu}, {Ma}, {Madsen}, {Mahn}, {Makino}, {Mancina}, {Maris}, {Maruyama}, {Mase}, {McElroy}, {McNally}, {Mead}, {Meagher}, {Medina}, {Meier}, {Meighen-Berger}, {Micallef}, {Mockler}, {Montaruli}, {Moore}, {Morse}, {Moulai}, {Naab}, {Nagai}, {Naumann}, {Necker}, {Nguyen}, {Niederhausen}, {Nisa}, {Nowicki}, {Nygren}, {Obertacke Pollmann}, {Oehler}, {Olivas}, {O'Sullivan}, {Pandya}, {Pankova}, {Park}, {Parker}, {Paudel}, {Paul}, {Perez de los Heros}, {Peters}, {Peterson}, {Philippen}, {Pieloth}, {Pieper}, {Pittermann}, {Pizzuto}, {Plum}, {Popovych}, {Porcelli}, {Prado Rodriguez}, {Price}, {Pries}, {Przybylski}, {Raab}, {Raissi}, {Rameez}, {Rawlins}, {Rea}, {Rehman}, {Reichherzer}, {Reimann}, {Renzi}, {Resconi}, {Reusch},
  {Rhode}, {Richman}, {Riedel}, {Roberts}, {Robertson}, {Roellinghoff}, {Rongen}, {Rott}, {Ruhe}, {Ryckbosch}, {Rysewyk Cantu}, {Safa}, {Saffer}, {Sanchez Herrera}, {Sandrock}, {Sandroos}, {Sarkar}, {Sarkar}, {Satalecka}, {Scharf}, {Schaufel}, {Schieler}, {Schindler}, {Schlunder}, {Schmidt}, {Schneider}, {Schneider}, {Schr{\"o}der}, {Schumacher}, {Schwefer}, {Sclafani}, {Seckel}, {Seunarine}, {Sharma}, {Shefali}, {Silva}, {Skrzypek}, {Smithers}, {Snihur}, {Soedingrekso}, {Soldin}, {Spannfellner}, {Spiczak}, {Spiering}, {Stachurska}, {Stamatikos}, {Stanev}, {Stein}, {Stettner}, {Steuer}, {Stezelberger}, {Sturwald}, {Stuttard}, {Sullivan}, {Taboada}, {Tenholt}, {Ter-Antonyan}, {Tilav}, {Tischbein}, {Tollefson}, {Tomankova}, {T{\"o}nnis}, {Toscano}, {Tosi}, {Trettin}, {Tselengidou}, {Tung}, {Turcati}, {Turcotte}, {Turley}, {Twagirayezu}, {Ty}, {Unland Elorrieta}, {Valtonen-Mattila}, {Vandenbroucke}, {van Eijndhoven}, {Vannerom}, {van Santen}, {Verpoest}, {Vraeghe}, {Walck}, {Watson}, {Weaver}, {Weigel},
  {Weindl}, {Weiss}, {Weldert}, {Wendt}, {Werthebach}, {Weyrauch}, {Whitehorn}, {Wiebusch}, {Williams}, {Wolf}, {Woschnagg}, {Wrede}, {Wulff}, {Xu}, {Xu}, {Yanez}, {Yoshida}, {Yu}, {Yuan}, \& {Zhang}}]{IceCube2022icrc.confE1142G}
{Goswami}, S., {Privon}, G.~C., {Santander}, M., {et~al.} 2022, in 37th International Cosmic Ray Conference, 1142

\bibitem[{{Greenhill} {et~al.}(1996){Greenhill}, {Gwinn}, {Antonucci}, \& {Barvainis}}]{Greenhill1996ApJ...472L..21G}
{Greenhill}, L.~J., {Gwinn}, C.~R., {Antonucci}, R., \& {Barvainis}, R. 1996, \apjl, 472, L21

\bibitem[{{Guti{\'e}rrez} {et~al.}(2021){Guti{\'e}rrez}, {Vieyro}, \& {Romero}}]{Gutierrez2021A&A...649A..87G}
{Guti{\'e}rrez}, E.~M., {Vieyro}, F.~L., \& {Romero}, G.~E. 2021, \aap, 649, A87

\bibitem[{{Haardt} \& {Maraschi}(1991)}]{Haardt1991}
{Haardt}, F. \& {Maraschi}, L. 1991, \apjl, 380, L51

\bibitem[{{Hagen} \& {Done}(2023)}]{Hagen2023MNRAS.525.3455H}
{Hagen}, S. \& {Done}, C. 2023, \mnras, 525, 3455

\bibitem[{{Hogg} \& {Reynolds}(2018)}]{Hogg2018ApJ...861...24H}
{Hogg}, J.~D. \& {Reynolds}, C.~S. 2018, \apj, 861, 24

\bibitem[{{Hooper} \& {Plant}(2023)}]{Hooper2023arXiv230506375H}
{Hooper}, D. \& {Plant}, K. 2023, arXiv e-prints, arXiv:2305.06375

\bibitem[{{Hoshino}(2012)}]{Hoshino2012PhRvL.108m5003H}
{Hoshino}, M. 2012, \prl, 108, 135003

\bibitem[{{Hur{\'e}}(2002)}]{Hure2002A&A...395L..21H}
{Hur{\'e}}, J.~M. 2002, \aap, 395, L21

\bibitem[{{IceCube Collaboration} {et~al.}(2022){IceCube Collaboration}, {Abbasi}, {Ackermann}, {Adams}, {Aguilar}, {Ahlers}, {Ahrens}, {Alameddine}, {Alispach}, {Alves}, {Amin}, {Andeen}, {Anderson}, {Anton}, {Arg{\"u}elles}, {Ashida}, {Axani}, {Bai}, {Balagopal}, {Barbano}, {Barwick}, {Bastian}, {Basu}, {Baur}, {Bay}, {Beatty}, {Becker}, {Becker Tjus}, {Bellenghi}, {Benzvi}, {Berley}, {Bernardini}, {Besson}, {Binder}, {Bindig}, {Blaufuss}, {Blot}, {Boddenberg}, {Bontempo}, {Borowka}, {B{\"o}ser}, {Botner}, {B{\"o}ttcher}, {Bourbeau}, {Bradascio}, {Braun}, {Brinson}, {Bron}, {Brostean-Kaiser}, {Browne}, {Burgman}, {Burley}, {Busse}, {Campana}, {Carnie-Bronca}, {Chen}, {Chen}, {Chirkin}, {Choi}, {Clark}, {Clark}, {Classen}, {Coleman}, {Collin}, {Conrad}, {Coppin}, {Correa}, {Cowen}, {Cross}, {Dappen}, {Dave}, {de Clercq}, {Delaunay}, {Delgado L{\'o}pez}, {Dembinski}, {Deoskar}, {Desai}, {Desiati}, {de Vries}, {de Wasseige}, {de With}, {Deyoung}, {Diaz}, {D{\'\i}az-V{\'e}lez}, {Dittmer}, {Dujmovic}, {Dunkman},
  {Duvernois}, {Dvorak}, {Ehrhardt}, {Eller}, {Engel}, {Erpenbeck}, {Evans}, {Evenson}, {Fan}, {Fazely}, {Fedynitch}, {Feigl}, {Fiedlschuster}, {Fienberg}, {Filimonov}, {Finley}, {Fischer}, {Fox}, {Franckowiak}, {Friedman}, {Fritz}, {F{\"u}rst}, {Gaisser}, {Gallagher}, {Ganster}, {Garcia}, {Garrappa}, {Gerhardt}, {Ghadimi}, {Glaser}, {Glauch}, {Gl{\"u}senkamp}, {Goldschmidt}, {Gonzalez}, {Goswami}, {Grant}, {Gr{\'e}goire}, {Griswold}, {G{\"u}nther}, {Gutjahr}, {Haack}, {Hallgren}, {Halliday}, {Halve}, {Halzen}, {Hanson}, {Hardin}, {Harnisch}, {Haungs}, {Hebecker}, {Helbing}, {Henningsen}, {Hettinger}, {Hickford}, {Hignight}, {Hill}, {Hill}, {Hoffman}, {Hoffmann}, {Hokanson-Fasig}, {Hoshina}, {Huang}, {Huber}, {Huber}, {Hultqvist}, {H{\"u}nnefeld}, {Hussain}, {Hymon}, {in}, {Iovine}, {Ishihara}, {Jansson}, {Japaridze}, {Jeong}, {Jin}, {Jones}, {Kang}, {Kang}, {Kang}, {Kappes}, {Kappesser}, {Kardum}, {Karg}, {Karl}, {Karle}, {Katz}, {Kauer}, {Kellermann}, {Kelley}, {Kheirandish}, {Kin}, {Kintscher}, {Kiryluk},
  {Klein}, {Koirala}, {Kolanoski}, {Kontrimas}, {K{\"o}pke}, {Kopper}, {Kopper}, {Koskinen}, {Koundal}, {Kovacevich}, {Kowalski}, {Kozynets}, {Kun}, {Kurahashi}, {Lad}, {Lagunas Gualda}, {Lanfranchi}, {Larson}, {Lauber}, {Lazar}, {Lee}, {Leonard}, {Leszczy{\'n}ska}, {Li}, {Lincetto}, {Liu}, {Liubarska}, {Lohfink}, {Lozano Mariscal}, {Lu}, {Lucarelli}, {Ludwig}, {Luszczak}, {Lyu}, {Ma}, {Madsen}, {Mahn}, {Makino}, {Mancina}, {Mari{\c{s}}}, {Martinez-Soler}, {Maruyama}, {Mase}, {McElroy}, {McNally}, {Mead}, {Meagher}, {Mechbal}, {Medina}, {Meier}, {Meighen-Berger}, {Micallef}, {Mockler}, {Montaruli}, {Moore}, {Morse}, {Moulai}, {Naab}, {Nagai}, {Nahnhauer}, {Naumann}, {Necker}, {Nguyen}, {Niederhausen}, {Nisa}, {Nowicki}, {Nygren}, {Obertack}, {Pollmann}, {Oehler}, {Oeyen}, {Olivas}, {O'Sullivan}, {Pandya}, {Pankova}, {Park}, {Parker}, {Paudel}, {Paul}, {P{\'e}rez de Los Heros}, {Peters}, {Peterson}, {Philippen}, {Pieper}, {Pittermann}, {Pizzuto}, {Plum}, {Popovych}, {Porcelli}, {Prado Rodriguez}, {Price},
  {Pries}, {Przybylski}, {Rack-Helleis}, {Raissi}, {Rameez}, {Rawlins}, {Rea}, {Rehman}, {Reichherzer}, {Reimann}, {Renzi}, {Resconi}, {Reusch}, {Rhode}, {Richman}, {Riedel}, {Roberts}, {Robertson}, {Roellinghoff}, {Rongen}, {Rott}, {Ruhe}, {Ryckbosch}, {Rysewyk Cantu}, {Safa}, {Saffer}, {Sanchez Herrera}, {Sandrock}, {Sandroos}, {Santander}, {Sarkar}, {Sarkar}, {Satalecka}, {Schaufel}, {Schieler}, {Schindler}, {Schmidt}, {Schneider}, {Schneider}, {Schr{\"o}der}, {Schumacher}, {Schwefer}, {Sclafani}, {Seckel}, {Seunarine}, {Sharma}, {Shefali}, {Silva}, {Skrzypek}, {Smithers}, {Snihur}, {Soedingrekso}, {Soldin}, {Spannfellner}, {Spiczak}, {Spiering}, {Stachurska}, {Stamatikos}, {Stanev}, {Stein}, {Stettner}, {Steuer}, {Stezelberger}, {Stokstad}, {St{\"u}rwald}, {Stuttard}, {Sullivan}, {Taboada}, {Ter-Antonyan}, {Tilav}, {Tischbein}, {Tollefson}, {T{\"o}nnis}, {Toscano}, {Tosi}, {Trettin}, {Tselengidou}, {Tung}, {Turcati}, {Turcotte}, {Turley}, {Twagirayezu}, {Ty}, {Unland Elorrieta}, {Valtonen-Mattila},
  {Vandenbroucke}, {van Eijndhoven}, {Vannerom}, {van Santen}, {Verpoest}, {Walck}, {Watson}, {Weaver}, {Weigel}, {Weindl}, {Weiss}, {Weldert}, {Wendt}, {Werthebach}, {Weyrauch}, {Whitehorn}, {Wiebusch}, {Williams}, {Wolf}, {Woschnagg}, {Wrede}, {Wulff}, {Xu}, {Yanez}, {Yoshida}, {Yu}, {Yuan}, {Zhangan}, \& {Zhelnin}}]{IceCube2022Sci...378..538I}
{IceCube Collaboration}, {Abbasi}, R., {Ackermann}, M., {et~al.} 2022, Science, 378, 538

\bibitem[{{Inoue} {et~al.}(2022){Inoue}, {Cerruti}, {Murase}, \& {Liu}}]{Inoue2022arXiv220702097I}
{Inoue}, S., {Cerruti}, M., {Murase}, K., \& {Liu}, R.-Y. 2022, arXiv e-prints, arXiv:2207.02097

\bibitem[{{Inoue} \& {Doi}(2018)}]{Inoue2018ApJ...869..114I}
{Inoue}, Y. \& {Doi}, A. 2018, \apj, 869, 114

\bibitem[{{Inoue} {et~al.}(2020){Inoue}, {Khangulyan}, \& {Doi}}]{Inoue2020ApJ...891L..33I}
{Inoue}, Y., {Khangulyan}, D., \& {Doi}, A. 2020, \apjl, 891, L33

\bibitem[{{Inoue} {et~al.}(2019){Inoue}, {Khangulyan}, {Inoue}, \& {Doi}}]{Inoue2019ApJ...880...40I}
{Inoue}, Y., {Khangulyan}, D., {Inoue}, S., \& {Doi}, A. 2019, \apj, 880, 40

\bibitem[{{Jin} {et~al.}(2012){Jin}, {Ward}, {Done}, \& {Gelbord}}]{Jin2012}
{Jin}, C., {Ward}, M., {Done}, C., \& {Gelbord}, J. 2012, \mnras, 420, 1825

\bibitem[{{Kalashev} {et~al.}(2015){Kalashev}, {Semikoz}, \& {Tkachev}}]{Kalashev2015}
{Kalashev}, O., {Semikoz}, D., \& {Tkachev}, I. 2015, Soviet Journal of Experimental and Theoretical Physics, 120, 541

\bibitem[{{Kato} {et~al.}(2008){Kato}, {Fukue}, \& {Mineshige}}]{Kato2008}
{Kato}, S., {Fukue}, J., \& {Mineshige}, S. 2008, {Black-Hole Accretion Disks --- Towards a New Paradigm ---} (Kyoto University Press (Kyoto, Japan))

\bibitem[{{Katz}(1976)}]{Katz1976}
{Katz}, J.~I. 1976, \apj, 206, 910

\bibitem[{{Kawabata} \& {Mineshige}(2010)}]{Kawabata2010PASJ...62..621K}
{Kawabata}, R. \& {Mineshige}, S. 2010, \pasj, 62, 621

\bibitem[{{Kawamuro} {et~al.}(2022){Kawamuro}, {Ricci}, {Imanishi}, {Mushotzky}, {Izumi}, {Ricci}, {Bauer}, {Koss}, {Trakhtenbrot}, {Ichikawa}, {Rojas}, {Smith}, {Shimizu}, {Oh}, {den Brok}, {Baba}, {Balokovi{\'c}}, {Chang}, {Kakkad}, {Pfeifle}, {Privon}, {Temple}, {Ueda}, {Harrison}, {Powell}, {Stern}, {Urry}, \& {Sanders}}]{Kawamuro2022ApJ...938...87K}
{Kawamuro}, T., {Ricci}, C., {Imanishi}, M., {et~al.} 2022, \apj, 938, 87

\bibitem[{{Kawamuro} {et~al.}(2023){Kawamuro}, {Ricci}, {Mushotzky}, {Imanishi}, {Bauer}, {Ricci}, {Koss}, {Privon}, {Trakhtenbrot}, {Izumi}, {Ichikawa}, {Rojas}, {Smith}, {Shimizu}, {Oh}, {den Brok}, {Baba}, {Balokovic}, {Chang}, {Kakkad}, {Pfeifle}, {Temple}, {Ueda}, {Harrison}, {Powell}, {Stern}, {Urry}, \& {Sanders}}]{Kawamuro2023arXiv230902776K}
{Kawamuro}, T., {Ricci}, C., {Mushotzky}, R.~F., {et~al.} 2023, arXiv e-prints, arXiv:2309.02776

\bibitem[{{Kawanaka} {et~al.}(2008){Kawanaka}, {Kato}, \& {Mineshige}}]{Kawanaka2008PASJ...60..399K}
{Kawanaka}, N., {Kato}, Y., \& {Mineshige}, S. 2008, \pasj, 60, 399

\bibitem[{{Kazanas} \& {Ellison}(1986)}]{Kazanas1986}
{Kazanas}, D. \& {Ellison}, D.~C. 1986, \apj, 304, 178

\bibitem[{{Kheirandish} {et~al.}(2021){Kheirandish}, {Murase}, \& {Kimura}}]{Kheirandish2021ApJ...922...45K}
{Kheirandish}, A., {Murase}, K., \& {Kimura}, S.~S. 2021, \apj, 922, 45

\bibitem[{{Kimura} {et~al.}(2015){Kimura}, {Murase}, \& {Toma}}]{Kimura2015ApJ...806..159K}
{Kimura}, S.~S., {Murase}, K., \& {Toma}, K. 2015, \apj, 806, 159

\bibitem[{{Kimura} {et~al.}(2019){Kimura}, {Tomida}, \& {Murase}}]{Kimura2019MNRAS.485..163K}
{Kimura}, S.~S., {Tomida}, K., \& {Murase}, K. 2019, \mnras, 485, 163

\bibitem[{{Koss} {et~al.}(2017){Koss}, {Trakhtenbrot}, {Ricci}, {Lamperti}, {Oh}, {Berney}, {Schawinski}, {Balokovi{\'c}}, {Baronchelli}, {Crenshaw}, {Fischer}, {Gehrels}, {Harrison}, {Hashimoto}, {Hogg}, {Ichikawa}, {Masetti}, {Mushotzky}, {Sartori}, {Stern}, {Treister}, {Ueda}, {Veilleux}, \& {Winter}}]{Koss2017ApJ...850...74K}
{Koss}, M., {Trakhtenbrot}, B., {Ricci}, C., {et~al.} 2017, \apj, 850, 74

\bibitem[{{Kubota} \& {Done}(2018)}]{Kubota2018MNRAS.480.1247K}
{Kubota}, A. \& {Done}, C. 2018, \mnras, 480, 1247

\bibitem[{Landau \& Lifshitz(1987)}]{Landau1987Fluid}
Landau, L.~D. \& Lifshitz, E.~M. 1987, Fluid Mechanics, Second Edition: Volume 6 (Course of Theoretical Physics), 2nd edn., Course of theoretical physics / by L. D. Landau and E. M. Lifshitz, Vol. 6 (Butterworth-Heinemann)

\bibitem[{{Liska} {et~al.}(2022){Liska}, {Musoke}, {Tchekhovskoy}, {Porth}, \& {Beloborodov}}]{Liska2022ApJ...935L...1L}
{Liska}, M.~T.~P., {Musoke}, G., {Tchekhovskoy}, A., {Porth}, O., \& {Beloborodov}, A.~M. 2022, \apjl, 935, L1

\bibitem[{{Liu} {et~al.}(2002{\natexlab{a}}){Liu}, {Mineshige}, {Meyer}, {Meyer-Hofmeister}, \& {Kawaguchi}}]{Liu2002ApJ...575..117L}
{Liu}, B.~F., {Mineshige}, S., {Meyer}, F., {Meyer-Hofmeister}, E., \& {Kawaguchi}, T. 2002{\natexlab{a}}, \apj, 575, 117

\bibitem[{{Liu} {et~al.}(2002{\natexlab{b}}){Liu}, {Mineshige}, \& {Shibata}}]{Liu2002}
{Liu}, B.~F., {Mineshige}, S., \& {Shibata}, K. 2002{\natexlab{b}}, \apj, 572, L173

\bibitem[{{Lodato} \& {Bertin}(2003)}]{Lodato2003A&A...398..517L}
{Lodato}, G. \& {Bertin}, G. 2003, \aap, 398, 517

\bibitem[{{Lynn} {et~al.}(2014){Lynn}, {Quataert}, {Chandran}, \& {Parrish}}]{Lynn2014ApJ...791...71L}
{Lynn}, J.~W., {Quataert}, E., {Chandran}, B. D.~G., \& {Parrish}, I.~J. 2014, \apj, 791, 71

\bibitem[{{Mahadevan} \& {Quataert}(1997)}]{Mahadevan1997ApJ...490..605M}
{Mahadevan}, R. \& {Quataert}, E. 1997, \apj, 490, 605

\bibitem[{{Manmoto} {et~al.}(1997){Manmoto}, {Mineshige}, \& {Kusunose}}]{Manmoto1997ApJ...489..791M}
{Manmoto}, T., {Mineshige}, S., \& {Kusunose}, M. 1997, \apj, 489, 791

\bibitem[{{Marinucci} {et~al.}(2016){Marinucci}, {Bianchi}, {Matt}, {Alexander}, {Balokovi{\'c}}, {Bauer}, {Brandt}, {Gandhi}, {Guainazzi}, {Harrison}, {Iwasawa}, {Koss}, {Madsen}, {Nicastro}, {Puccetti}, {Ricci}, {Stern}, \& {Walton}}]{Marinucci2016MNRAS.456L..94M}
{Marinucci}, A., {Bianchi}, S., {Matt}, G., {et~al.} 2016, \mnras, 456, L94

\bibitem[{{Matsumoto} {et~al.}(1988){Matsumoto}, {Horiuchi}, {Shibata}, \& {Hanawa}}]{Matsumoto1988PASJ...40..171M}
{Matsumoto}, R., {Horiuchi}, T., {Shibata}, K., \& {Hanawa}, T. 1988, \pasj, 40, 171

\bibitem[{{Mbarek} {et~al.}(2023){Mbarek}, {Philippov}, {Chernoglazov}, {Levinson}, \& {Mushotzky}}]{Mbarek2023arXiv231015222M}
{Mbarek}, R., {Philippov}, A., {Chernoglazov}, A., {Levinson}, A., \& {Mushotzky}, R. 2023, arXiv e-prints, arXiv:2310.15222

\bibitem[{{Michiyama} {et~al.}(2023){Michiyama}, {Inoue}, \& {Doi}}]{Michiyama2023PASJ..tmp...58M}
{Michiyama}, T., {Inoue}, Y., \& {Doi}, A. 2023, \pasj, 75, 874

\bibitem[{{Minezaki} \& {Matsushita}(2015)}]{Minezaki2015ApJ...802...98M}
{Minezaki}, T. \& {Matsushita}, K. 2015, \apj, 802, 98

\bibitem[{{Mizumoto} {et~al.}(2019){Mizumoto}, {Izumi}, \& {Kohno}}]{Mizumoto2019ApJ...871..156M}
{Mizumoto}, M., {Izumi}, T., \& {Kohno}, K. 2019, \apj, 871, 156

\bibitem[{{Morgan} {et~al.}(2012){Morgan}, {Hainline}, {Chen}, {Tewes}, {Kochanek}, {Dai}, {Kozlowski}, {Blackburne}, {Mosquera}, {Chartas}, {Courbin}, \& {Meylan}}]{Morgan2012}
{Morgan}, C.~W., {Hainline}, L.~J., {Chen}, B., {et~al.} 2012, \apj, 756, 52

\bibitem[{{Morishima} {et~al.}(2023){Morishima}, {Sudou}, {Yamauchi}, {Taniguchi}, \& {Nakai}}]{Morishima2023PASJ...75...71M}
{Morishima}, Y., {Sudou}, H., {Yamauchi}, A., {Taniguchi}, Y., \& {Nakai}, N. 2023, \pasj, 75, 71

\bibitem[{{M{\"u}ller} {et~al.}(2022){M{\"u}ller}, {Naddaf}, {Zaja{\v{c}}ek}, {Czerny}, {Araudo}, \& {Karas}}]{Muller2022ApJ...931...39M}
{M{\"u}ller}, A.~L., {Naddaf}, M.-H., {Zaja{\v{c}}ek}, M., {et~al.} 2022, \apj, 931, 39

\bibitem[{{M{\"u}ller} \& {Romero}(2020)}]{Muller2020A&A...636A..92M}
{M{\"u}ller}, A.~L. \& {Romero}, G.~E. 2020, \aap, 636, A92

\bibitem[{{Murase}(2022)}]{Murase2022ApJ...941L..17M}
{Murase}, K. 2022, \apjl, 941, L17

\bibitem[{{Murase} {et~al.}(2020){Murase}, {Kimura}, \& {M{\'e}sz{\'a}ros}}]{Murase2020PhRvL.125a1101M}
{Murase}, K., {Kimura}, S.~S., \& {M{\'e}sz{\'a}ros}, P. 2020, \prl, 125, 011101

\bibitem[{{Narayan} \& {Yi}(1994)}]{Narayan1994ApJ...428L..13N}
{Narayan}, R. \& {Yi}, I. 1994, \apjl, 428, L13

\bibitem[{{Neronov} {et~al.}(2023){Neronov}, {Savchenko}, \& {Semikoz}}]{Neronov2023arXiv230609018N}
{Neronov}, A., {Savchenko}, D., \& {Semikoz}, D.~V. 2023, arXiv e-prints, arXiv:2306.09018

\bibitem[{{Nishizuka} \& {Shibata}(2013)}]{Nishizuka2013}
{Nishizuka}, N. \& {Shibata}, K. 2013, Physical Review Letters, 110, 051101

\bibitem[{{Oh} {et~al.}(2018){Oh}, {Koss}, {Markwardt}, {Schawinski}, {Baumgartner}, {Barthelmy}, {Cenko}, {Gehrels}, {Mushotzky}, {Petulante}, {Ricci}, {Lien}, \& {Trakhtenbrot}}]{Oh2018ApJS..235....4O}
{Oh}, K., {Koss}, M., {Markwardt}, C.~B., {et~al.} 2018, \apjs, 235, 4

\bibitem[{{Pal} {et~al.}(2023){Pal}, {Anju}, {Sreehari}, {Rameshan}, {Stalin}, {Ricci}, \& {Marchesi}}]{Pal2023arXiv231018196P}
{Pal}, I., {Anju}, A., {Sreehari}, H., {et~al.} 2023, arXiv e-prints, arXiv:2310.18196

\bibitem[{{Panessa} {et~al.}(2022){Panessa}, {P{\'e}rez-Torres}, {Hern{\'a}ndez-Garc{\'\i}a}, {Casella}, {Giroletti}, {Orienti}, {Baldi}, {Bassani}, {Fiocchi}, {La Franca}, {Malizia}, {McHardy}, {Nicastro}, {Piro}, {Vincentelli}, {Williams}, \& {Ubertini}}]{Panessa2022MNRAS.510..718P}
{Panessa}, F., {P{\'e}rez-Torres}, M., {Hern{\'a}ndez-Garc{\'\i}a}, L., {et~al.} 2022, \mnras, 510, 718

\bibitem[{{Parker}(1955)}]{Parker1955ApJ...121..491P}
{Parker}, E.~N. 1955, \apj, 121, 491

\bibitem[{{Parker}(1966)}]{Parker1966ApJ...145..811P}
{Parker}, E.~N. 1966, \apj, 145, 811

\bibitem[{{Peretti} {et~al.}(2023){Peretti}, {Peron}, {Tombesi}, {Lamastra}, {Ahlers}, \& {Saturni}}]{Peretti2023arXiv230303298P}
{Peretti}, E., {Peron}, G., {Tombesi}, F., {et~al.} 2023, arXiv e-prints, arXiv:2303.03298

\bibitem[{{Petrucci} {et~al.}(2023){Petrucci}, {Pi{\'e}tu}, {Behar}, {Clavel}, {Bianchi}, {Henri}, {Barnier}, {Chen}, {Ferreira}, {Malzac}, {Belmont}, {Corbel}, \& {Coriat}}]{Petrucci2023arXiv230901804P}
{Petrucci}, P.~O., {Pi{\'e}tu}, V., {Behar}, E., {et~al.} 2023, arXiv e-prints, arXiv:2309.01804

\bibitem[{{Porth} {et~al.}(2021){Porth}, {Mizuno}, {Younsi}, \& {Fromm}}]{Porth2021MNRAS.502.2023P}
{Porth}, O., {Mizuno}, Y., {Younsi}, Z., \& {Fromm}, C.~M. 2021, \mnras, 502, 2023

\bibitem[{{Pozdniakov} {et~al.}(1977){Pozdniakov}, {Sobol}, \& {Siuniaev}}]{Pozdniakov1977}
{Pozdniakov}, L.~A., {Sobol}, I.~M., \& {Siuniaev}, R.~A. 1977, \sovast, 21, 708

\bibitem[{{Raginski} \& {Laor}(2016)}]{Raginski2016MNRAS.459.2082R}
{Raginski}, I. \& {Laor}, A. 2016, \mnras, 459, 2082

\bibitem[{{Reynolds}(2021)}]{Reynolds2021ARA&A..59..117R}
{Reynolds}, C.~S. 2021, \araa, 59, 117

\bibitem[{{Ricci} {et~al.}(2023){Ricci}, {Chang}, {Kawamuro}, {Privon}, {Mushotzky}, {Trakhtenbrot}, {Laor}, {Koss}, {Smith}, {Gupta}, {Dimopoulos}, {Aalto}, \& {Ros}}]{Ricci2023ApJ...952L..28R}
{Ricci}, C., {Chang}, C.-S., {Kawamuro}, T., {et~al.} 2023, \apjl, 952, L28

\bibitem[{{Ricci} {et~al.}(2018){Ricci}, {Ho}, {Fabian}, {Trakhtenbrot}, {Koss}, {Ueda}, {Lohfink}, {Shimizu}, {Bauer}, {Mushotzky}, {Schawinski}, {Paltani}, {Lamperti}, {Treister}, \& {Oh}}]{Ricci2018MNRAS.480.1819R}
{Ricci}, C., {Ho}, L.~C., {Fabian}, A.~C., {et~al.} 2018, \mnras, 480, 1819

\bibitem[{{Ripperda} {et~al.}(2022){Ripperda}, {Liska}, {Chatterjee}, {Musoke}, {Philippov}, {Markoff}, {Tchekhovskoy}, \& {Younsi}}]{Ripperda2022ApJ...924L..32R}
{Ripperda}, B., {Liska}, M., {Chatterjee}, K., {et~al.} 2022, \apjl, 924, L32

\bibitem[{{Shibata} {et~al.}(2013){Shibata}, {Isobe}, {Hillier}, {Choudhuri}, {Maehara}, {Ishii}, {Shibayama}, {Notsu}, {Notsu}, {Nagao}, {Honda}, \& {Nogami}}]{Shibata2013PASJ...65...49S}
{Shibata}, K., {Isobe}, H., {Hillier}, A., {et~al.} 2013, \pasj, 65, 49

\bibitem[{{Shimizu} {et~al.}(2016){Shimizu}, {Mel{\'e}ndez}, {Mushotzky}, {Koss}, {Barger}, \& {Cowie}}]{Shimizu2016MNRAS.456.3335S}
{Shimizu}, T.~T., {Mel{\'e}ndez}, M., {Mushotzky}, R.~F., {et~al.} 2016, \mnras, 456, 3335

\bibitem[{{Sikora} \& {Begelman}(2013)}]{Sikora2013ApJ...764L..24S}
{Sikora}, M. \& {Begelman}, M.~C. 2013, \apjl, 764, L24

\bibitem[{{Sikora} {et~al.}(1987){Sikora}, {Kirk}, {Begelman}, \& {Schneider}}]{Sikora1987}
{Sikora}, M., {Kirk}, J.~G., {Begelman}, M.~C., \& {Schneider}, P. 1987, \apjl, 320, L81

\bibitem[{{Sikora} {et~al.}(2007){Sikora}, {Stawarz}, \& {Lasota}}]{Sikora2007ApJ...658..815S}
{Sikora}, M., {Stawarz}, {\L}., \& {Lasota}, J.-P. 2007, \apj, 658, 815

\bibitem[{{Singh} {et~al.}(2015){Singh}, {de Gouveia Dal Pino}, \& {Kadowaki}}]{Singh2015ApJ...799L..20S}
{Singh}, C.~B., {de Gouveia Dal Pino}, E.~M., \& {Kadowaki}, L.~H.~S. 2015, \apjl, 799, L20

\bibitem[{{Sotomayor} \& {Romero}(2022)}]{Sotomayor2022A&A...664A.178S}
{Sotomayor}, P. \& {Romero}, G.~E. 2022, \aap, 664, A178

\bibitem[{{Spitzer}(1962)}]{Spitzer1962}
{Spitzer}, L. 1962, {Physics of Fully Ionized Gases} (New York: Interscience)

\bibitem[{{Sridhar} {et~al.}(2022){Sridhar}, {Metzger}, \& {Fang}}]{Sridhar2022arXiv221211236S}
{Sridhar}, N., {Metzger}, B.~D., \& {Fang}, K. 2022, arXiv e-prints, arXiv:2212.11236

\bibitem[{{Stecker} {et~al.}(1992){Stecker}, {Done}, {Salamon}, \& {Sommers}}]{Stecker1992}
{Stecker}, F.~W., {Done}, C., {Salamon}, M.~H., \& {Sommers}, P. 1992, Physical Review Letters, 69, 2738

\bibitem[{{Stepney}(1983)}]{Stepney1983}
{Stepney}, S. 1983, \mnras, 202, 467

\bibitem[{{Sun} \& {Bai}(2021)}]{Sun2021MNRAS.506.1128S}
{Sun}, X. \& {Bai}, X.-N. 2021, \mnras, 506, 1128

\bibitem[{{Sunyaev} \& {Titarchuk}(1980)}]{Sunyaev1980}
{Sunyaev}, R.~A. \& {Titarchuk}, L.~G. 1980, \aap, 500, 167

\bibitem[{{Tagawa} {et~al.}(2023){Tagawa}, {Kimura}, \& {Haiman}}]{Tagawa2023ApJ...955...23T}
{Tagawa}, H., {Kimura}, S.~S., \& {Haiman}, Z. 2023, \apj, 955, 23

\bibitem[{{Takasao} {et~al.}(2018){Takasao}, {Tomida}, {Iwasaki}, \& {Suzuki}}]{Takasao2018ApJ...857....4T}
{Takasao}, S., {Tomida}, K., {Iwasaki}, K., \& {Suzuki}, T.~K. 2018, \apj, 857, 4

\bibitem[{{Takasao} {et~al.}(2019){Takasao}, {Tomida}, {Iwasaki}, \& {Suzuki}}]{Takasao2019ApJ...878L..10T}
{Takasao}, S., {Tomida}, K., {Iwasaki}, K., \& {Suzuki}, T.~K. 2019, \apjl, 878, L10

\bibitem[{{Tanaka} {et~al.}(1995){Tanaka}, {Nandra}, {Fabian}, {Inoue}, {Otani}, {Dotani}, {Hayashida}, {Iwasawa}, {Kii}, {Kunieda}, {Makino}, \& {Matsuoka}}]{Tanaka1995Natur.375..659T}
{Tanaka}, Y., {Nandra}, K., {Fabian}, A.~C., {et~al.} 1995, \nat, 375, 659

\bibitem[{{Tanimoto} {et~al.}(2018){Tanimoto}, {Ueda}, {Kawamuro}, {Ricci}, {Awaki}, \& {Terashima}}]{Tanimoto2018ApJ...853..146T}
{Tanimoto}, A., {Ueda}, Y., {Kawamuro}, T., {et~al.} 2018, \apj, 853, 146

\bibitem[{{Tanimoto} {et~al.}(2022){Tanimoto}, {Ueda}, {Odaka}, {Yamada}, \& {Ricci}}]{Tanimoto2022ApJS..260...30T}
{Tanimoto}, A., {Ueda}, Y., {Odaka}, H., {Yamada}, S., \& {Ricci}, C. 2022, \apjs, 260, 30

\bibitem[{{Vasudevan} {et~al.}(2016){Vasudevan}, {Fabian}, {Reynolds}, {Aird}, {Dauser}, \& {Gallo}}]{Vasudevan2016MNRAS.458.2012V}
{Vasudevan}, R.~V., {Fabian}, A.~C., {Reynolds}, C.~S., {et~al.} 2016, \mnras, 458, 2012

\bibitem[{{Volonteri} {et~al.}(2013){Volonteri}, {Sikora}, {Lasota}, \& {Merloni}}]{Volonteri2013ApJ...775...94V}
{Volonteri}, M., {Sikora}, M., {Lasota}, J.~P., \& {Merloni}, A. 2013, \apj, 775, 94

\bibitem[{{Yuan} \& {Narayan}(2014)}]{Yuan2014}
{Yuan}, F. \& {Narayan}, R. 2014, \araa, 52, 529

\bibitem[{{Yuan} {et~al.}(2012){Yuan}, {Wu}, \& {Bu}}]{Yuan2012ApJ...761..129Y}
{Yuan}, F., {Wu}, M., \& {Bu}, D. 2012, \apj, 761, 129

\bibitem[{{Yuan} {et~al.}(2020){Yuan}, {Fausnaugh}, {Hoffmann}, {Macri}, {Peterson}, {Riess}, {Bentz}, {Brown}, {Bont{\`a}}, {Davies}, {Rosa}, {Ferrarese}, {Grier}, {Hicks}, {Onken}, {Pogge}, {Storchi-Bergmann}, \& {Vestergaard}}]{Yuan2020ApJ...902...26Y}
{Yuan}, W., {Fausnaugh}, M.~M., {Hoffmann}, S.~L., {et~al.} 2020, \apj, 902, 26

\bibitem[{{Zdziarski}(1986)}]{Zdziarski1986}
{Zdziarski}, A.~A. 1986, \apj, 305, 45

\bibitem[{{Zhdankin} {et~al.}(2018){Zhdankin}, {Uzdensky}, {Werner}, \& {Begelman}}]{Zhdankin2018ApJ...867L..18Z}
{Zhdankin}, V., {Uzdensky}, D.~A., {Werner}, G.~R., \& {Begelman}, M.~C. 2018, \apjl, 867, L18

\bibitem[{{Zhdankin} {et~al.}(2019){Zhdankin}, {Uzdensky}, {Werner}, \& {Begelman}}]{Zhdankin2019PhRvL.122e5101Z}
{Zhdankin}, V., {Uzdensky}, D.~A., {Werner}, G.~R., \& {Begelman}, M.~C. 2019, \prl, 122, 055101

\end{thebibliography}
\end{document}